\documentclass[apj]{emulateapj}

\def\ltsima{$\; \buildrel < \over \sim \;$}
\def\simlt{\lower.5ex\hbox{\ltsima}}
\def\gtsima{$\; \buildrel > \over \sim \;$}
\def\simgt{\lower.5ex\hbox{\gtsima}}
\def\kpc{{\rm\,kpc}}

\def\msun{{\rm\,M_\odot}}
\def\lsun{{\rm\,L_\odot}}

\def\pc{{\rm\,pc}}

\def\deg{^\circ}

\def\s{\ifmmode \widetilde \else \~\fi}
\def\={\overline}

\def\spose#1{\hbox to 0pt{#1\hss}}

\def\lta{\mathrel{\spose{\lower 3pt\hbox{$\mathchar"218$}}
     \raise 2.0pt\hbox{$\mathchar"13C$}}}
\def\gta{\mathrel{\spose{\lower 3pt\hbox{$\mathchar"218$}}
     \raise 2.0pt\hbox{$\mathchar"13E$}}}
\def\Dt{\spose{\raise 1.5ex\hbox{\hskip3pt$\mathchar"201$}}}    
\def\dt{\spose{\raise 1.0ex\hbox{\hskip2pt$\mathchar"201$}}}    

\def\dotsfill{\leaders\hbox to 1em{\hss.\hss}\hfill}

\def\Gyr{{\rm\,Gyr}}
\def\FeH{{\rm[Fe/H]}}

\shorttitle{Two new dwarf galaxies in the surroundings of M31 and M33}
\shortauthors{N. F. Martin et al.}

\begin{document}


\title{PAndAS' cubs: discovery of two new dwarf galaxies in the surroundings of the Andromeda and Triangulum galaxies.$^{12}$}


\author{Nicolas F. Martin$^1$, Alan W. McConnachie$^2$, Mike Irwin$^3$, Lawrence M. Widrow$^4$, Annette M. N. Ferguson$^5$, Rodrigo A. Ibata$^6$, John Dubinski$^7$, Arif Babul$^8$, Scott Chapman $^3$, Mark Fardal$^9$, Geraint F. Lewis$^{10}$, Julio Navarro$^8$ and R. Michael Rich$^{11}$}
\email{martin@mpia.de}

\altaffiltext{1}{Max-Planck-Institut f\"ur Astronomie, K\"onigstuhl 17, D-69117 Heidelberg, Germany}
\altaffiltext{2}{NRC Herzberg Institute for Astrophysics, 5071 West Saanich Road, Victoria, British Columbia, Canada, V9E 2E7}
\altaffiltext{3}{Institute of Astronomy, Madingley Road, Cambridge, CB3 0HA, U.K.}
\altaffiltext{4}{Department of Physics, Engineering Physics, and Astronomy, Queen's University, Kingston, ON K7L 3N6, Canada}
\altaffiltext{5}{Institute for Astronomy, University of Edinburgh, Royal Observatory, Blackford Hill, Edinburgh, EH9 3HJ, U.K.}
\altaffiltext{6}{Observatoire de Strasbourg, 11, rue de l'Universit\'e, F-67000, Strasbourg, France}
\altaffiltext{7}{Department of Astronomy and Astrophysics, University of Toronto, 60 St. George Street, Toronto, ON M5S 3H8, Canada}
\altaffiltext{8}{Department of Physics and Astronomy, University of Victoria, 3800 Finnerty Rd., Victoria, BC V8W 3P6, Canada}
\altaffiltext{9}{Astronomy Department, University of Massachusetts, Amherst, MA 01003, USA}
\altaffiltext{10}{Institute of Astronomy, School of Physics, University of Sydney, NSW 2006, Australia.}
\altaffiltext{11}{Physics and Astronomy Building, 430 Portola Plaza, Box 951547, Department of Physics and Astronomy, University of California, Los Angeles, CA 90095-1547, USA.}

\begin{abstract}
We present the discovery of two new dwarf galaxies, Andromeda~XXI and Andromeda~XXII, located in the surroundings of the Andromeda and Triangulum galaxies (M31 and M33). These discoveries stem from the first year data of the Pan-Andromeda Archaeological Survey (PAndAS), a photometric survey of the M31/M33 group conducted with the Megaprime/MegaCam wide-field camera mounted on the Canada-France-Hawaii Telescope. Both satellites appear as spatial overdensities of stars which, when plotted in a color-magnitude diagram, follow metal-poor, $\FeH=-1.8$, red giant branches at the distance of M31/M33. Andromeda~XXI is a moderately bright dwarf galaxy ($M_V=-9.9\pm0.6$), albeit with low surface brightness, emphasizing again that many relatively luminous M31 satellites still remain to be discovered. It is also a large satellite, with a half-light radius close to $1\kpc$, making it the fourth largest Local Group dwarf spheroidal galaxy after the recently discovered Andromeda~XIX, Andromeda~II and Sagittarius around the Milky Way, and supports the trend that M31 satellites are larger than their Milky Way counterparts. Andromeda~XXII is much fainter ($M_V=-6.5\pm0.8$) and lies a lot closer in projection to M33 than it does to M31 (42 vs. 224\kpc), suggesting that it could be the first Triangulum satellite to be discovered. Although this is a very exciting possibility in the context of a past interaction of M33 with M31 and the fate of its satellite system, a confirmation will have to await a good distance estimate to confirm its physical proximity to M33. Along with the dwarf galaxies found in previous surveys of the M31 surroundings, these two new satellites bring the number of dwarf spheroidal galaxies in this region to 20.
\end{abstract}


\section{Introduction}
\footnotetext[12]{Based on observations obtained with MegaPrime/MegaCam, a joint project of CFHT and CEA/DAPNIA, at
the Canada-France-Hawaii Telescope (CFHT) which is operated by the National Research Council (NRC) of Canada, the
Institut National des Science de l'Univers of the Centre National de la Recherche Scientifique (CNRS) of France, and the
University of Hawaii.}
Systematic deep photometric observations of the surroundings of the Andromeda galaxy have revolutionized our knowledge of its satellite system. Before 2004, 12 dwarf galaxies were known to be M31 companions, including only 6 dwarf spheroidal galaxies, while 12 new dwarf galaxies have been discovered to inhabit this region of the Local Group sky over the last five years. A dedicated scan of the Sloan Digital Sky Survey (SDSS), targeting a region within $2\deg$ of M31's major axis, unveiled the presence of Andromeda~IX and~X \citep{zucker04,zucker07}. In parallel, a contiguous mapping of the region within $\sim50\kpc$ of Andromeda, performed with the Wide Field Camera on the Isaac Newton Telescope, revealed the picture of a strikingly substructured inner stellar halo \citep{ferguson02}, and led to the discovery of Andromeda~XVII \citep{irwin08}.

Mapping the outer regions of M31's halo was the goal of an extension to this survey. Performed with the MegaPrime/MegaCam $1\,\textrm{deg}^2$ camera on the Canada-France-Hawaii Telescope (CFHT), it mapped a quarter of the M31 stellar halo, out to $\sim150\kpc$ \citep{ibata07}. Confirming the clumpy nature of Andromeda's stellar halo, the survey has also revealed the presence of numerous other dwarf galaxies: Andromeda~XI, XII, XIII, XV, XVI, XVIII, XIX and~XX \citep{martin06b,ibata07,mcconnachie08}. At about the same time, Andromeda~XIV was discovered serendipitously by \citet{majewski07}, just outside of the edge of the MegaCam survey.

Our understanding of these systems remains sparse, mainly due to the distance at which they reside, translating into difficult photometric observations that cannot easily reach much deeper than the horizontal branch at the distance of M31. Spectroscopic observations are also limited to the handful of bright red giant branch (RGB) stars that can be targeted in each system. Although progress is expected along these lines in the coming years, much can already be said of the generic properties of these new objects from the current survey data alone. Their absolute magnitude ranges from a very faint $M_V=-6.4$ for And~XII and And~XX \citep{martin06b,mcconnachie08} to a surprisingly bright $M_V\lta-9.7$ for And~XVIII \citep{mcconnachie08}, patently showing the incompleteness of the M31 satellite luminosity function at the faint end, in regions that have so far only been surveyed with photographic plates.

Given their relative faintness and significant sizes, these new systems are usually assumed to be dwarf spheroidal galaxies, in other words dwarf galaxies that are devoid of any significant amount of gas. This is currently consistent with the analysis of \textsc{Hi} surveys but the upper limits on their \textsc{Hi} content remains relatively high ($2-3\times10^5\msun$; \citealt{grcevich09}). Therefore, it cannot be entirely ruled out that some of these galaxies still contain non-negligeable amounts of gas.

Building upon the previous MegaCam survey, we have initiated the Pan-Andromeda Archaeological Survey (PAndAS), a Large Program using the CFHT MegaCam imager to map the entire stellar halo of M31 and M33 out to distances of $\sim 150$ and $\sim50\kpc$ respectively. 
Here, we report on the discovery of two new dwarf galaxies in the vicinity of the Andromeda and Triangulum galaxies based on the first year of PAndAS data. These systems, dubbed Andromeda~XXI and~XXII are sparse but unmistakable overdensities of stars that are also aligned along a RGB at the distance of M31. Section~2 of this paper briefly summarizes the PAndAS data while section~3 presents the new systems and details their properties. Section~4 briefly discusses the new discoveries and section~5 concludes the paper.

Throughout this paper, the distance moduli of M31 and M33 are assumed to be $(m-M)_0=24.47\pm0.07$ and $(m-M)_0=24.54\pm0.06$, or $783\pm25$ and $809\pm24\kpc$ respectively \citep{mcconnachie05}.

\section{The PAndAS survey}
PAndAS builds upon our previous CFHT/MegaCam surveys of M31 whose results are presented in \citet{martin06b}, \citet{ibata07} and \citet{mcconnachie08}. We use a similar observational set-up and refer the reader to these papers for a description of the observing strategy, data reduction and data quality. Maps showing early results and the complete survey area to date can be found in \citet{mcconnachie09} and led to the discovery of And~XXI and And~XXII. The survey consists of contiguous exposures, performed with the $1\times1$ deg$^2$ MegaPrime/MegaCam camera mounted on the CFHT. The camera is a mosaic of 36 $2048\times4612$ CCDs with a pixel size of 0.187\,arsec. Small gaps within the survey lead to a scientifically useable field-of-view of $0.96\times0.94\,\textrm{deg}^2$ for each of the current 250\,pointings of the survey.

Each field has been observed for 1350\,s in each of the MegaCam $g$ and $i$ filters, split into $3\times450$\,s dithered subexposures. Good seeing ($<0.8''$) ensures that the photometry reaches $g\sim25.5$ and $i\sim24.5$ with a signal-to-noise ratio of 10 and guarantees that the star/galaxy separation only degrades at magnitudes fainter than $g\sim25.0$ and $i\sim24.0$. Data were preprocessed (de-biased, flat-fielded and fringe corrected) by the Elixir system at CFHT that also determines the photometric zero point of the observations, and then processed using a version of the Cambridge Astronomical Survey Unit (CASU) photometry pipeline \citep{irwin01}, adapted to CFHT/MegaCam observations. The pipeline registers and stacks the images, and also generates catalogues with object morphological classification before creating band-merged $g$, $i$ products.

In the following, dereddened magnitudes ($g_0$ and $i_0$) have been determined from the \citet{schlegel98} $E(B-V)$ extinction maps, using the following correction coefficients: $g_0=g-3.793E(B-V)$ and $i_0=i-2.086E(B-V)$, listed in their Table~6.

\section{Andromeda~XXI and Andromeda~XXII}
\subsection{Discovery}

Two new dwarf galaxies were directly identified on matched-filter surface density maps of RGB candidate stars selected to have colors consistent with the average (generally) metal-poor locus of M31 dwarf spheroidals. Satellites in the very sparsely populated outer halo regions correspond to stellar overdensities that are striking enough for them to be easily spotted, even down to magnitudes of $M_V\sim-6.5$. This was the case for And~XII \citep{martin06b} and And~XX \citep{mcconnachie08} and also here with Andromeda~XXII. An automated search is certainly warranted to better quantify the detection limits of the survey in uncovering new dwarf galaxies. However, it is beyond the scope of this initial discovery paper and will be presented elsewhere.

\begin{figure}
\begin{center}
\includegraphics[width=\hsize,angle=270]{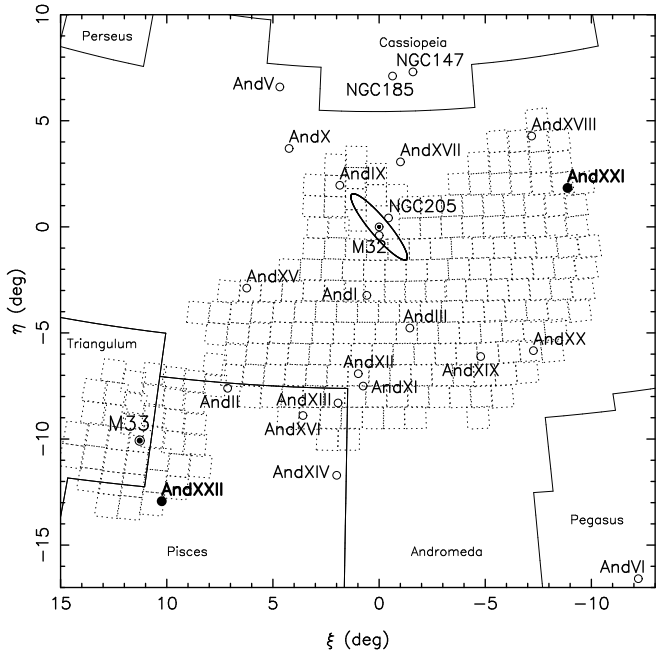}
\caption{\label{map} Map of dwarf galaxies located in the surroundings of M31 and M33. M31 is located at the center of the coordinate system and known satellites are represented by empty circles. The two new discoveries, And~XXI and And~XXII, are represented by filled circles. The current extent of the PAndAS survey is shown as dotted squares, with each square representing a single $1\times1\,\textrm{deg}^2$ MegaCam field. The ellipse represents the extent of M31's \textsc{Hi} disk. Constellation boundaries, taken from \citet{davenhall97}, are shown as thin lines.}
\end{center}
\end{figure}

The locations of the two new satellites are shown on the map of the PAndAS survey (Figure~\ref{map}). As with many previous detections, these two systems are noticeably close to the edge of the survey, suggesting that the survey limit of $150\kpc$ in projected distance from M31 does not represent the true extent of Andromeda's satellite system (see also Figure~2 of \citealt{mcconnachie09}). In the spirit of previous conventions, we dub these two satellites Andromeda~XXI and Andromeda~XXII (And~XXI and And~XXII; see the discussion below, in Appendix~\ref{naming_conventions}).

\begin{figure*}
\begin{center}
\includegraphics[width=0.44\hsize,angle=270]{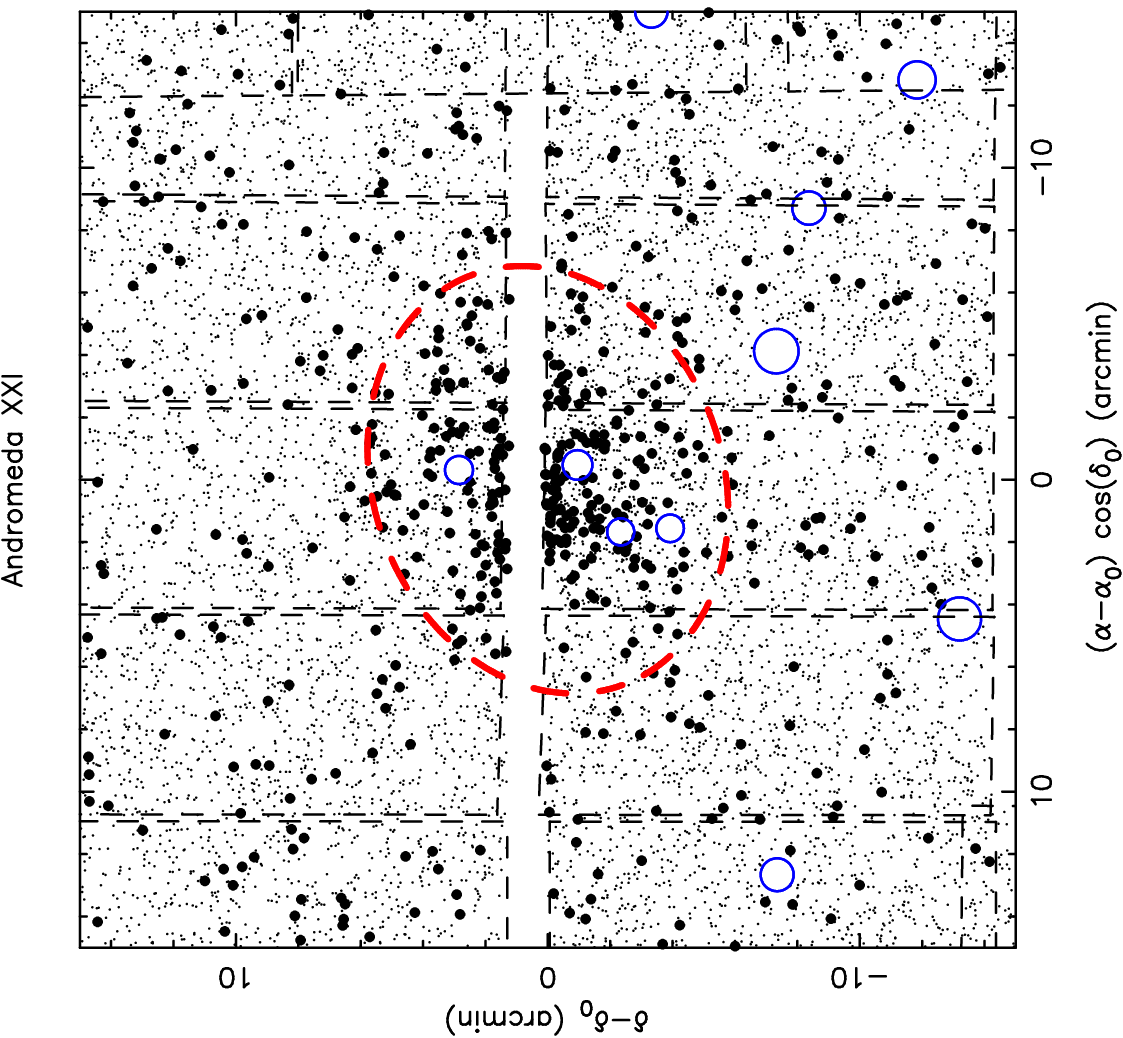}
\includegraphics[width=0.44\hsize,angle=270]{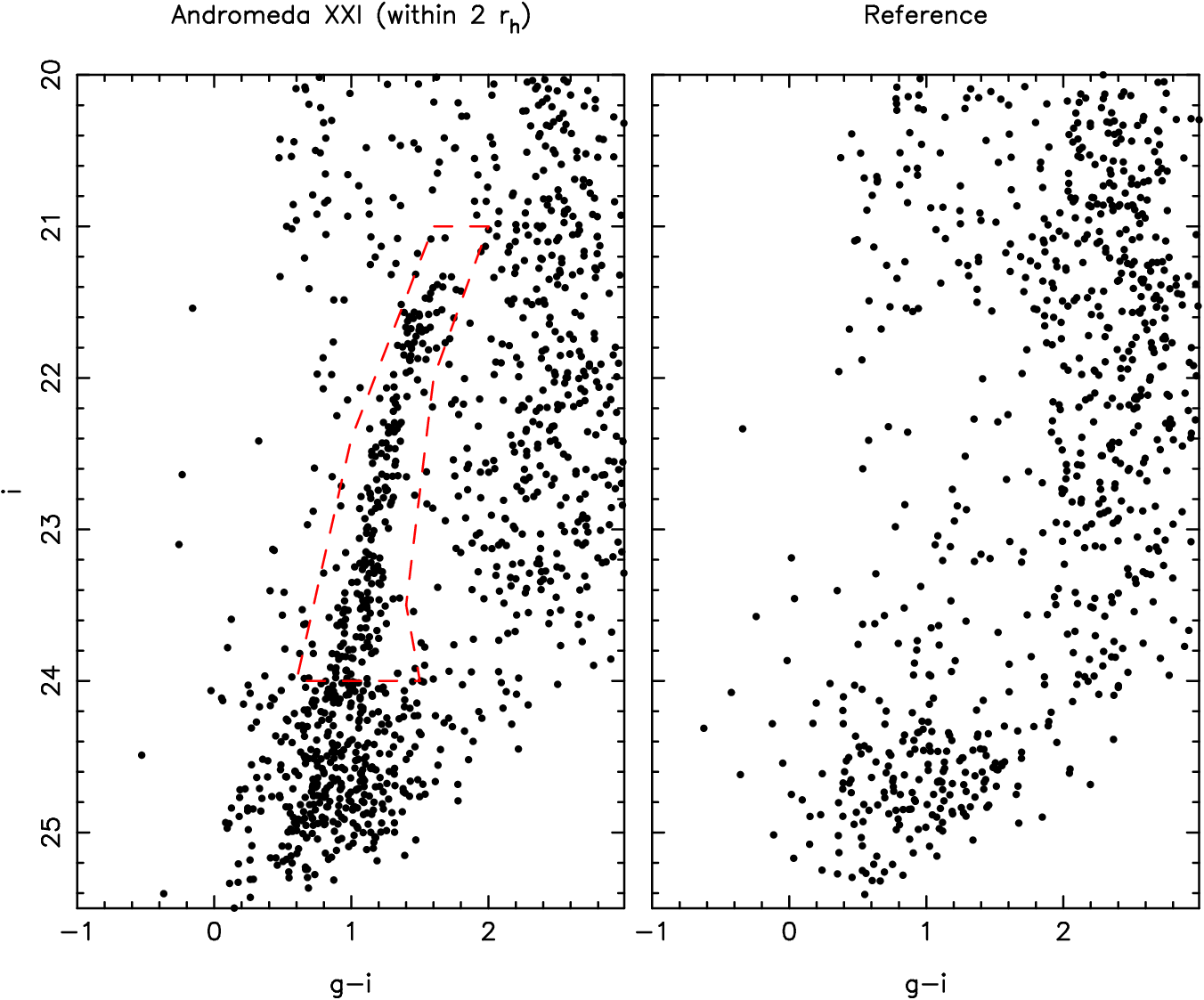}
\caption{\label{map_AndXXI}\emph{Left panel:} Spatial distribution of stellar sources around And~XXI. Small dots represent all stars in the PAndAS survey whereas large dots correspond to likely RGB stars of the dwarf galaxy, selected within the dashed box shown on the CMD of the middle panel. These stars are clearly clumped into an overdensity of stars. MegaCam CCDs are shown as dashed rectangles and white regions correspond to holes in-between CCDs or holes in the survey. Open circles correspond to regions that are lost to the survey due to the presence of saturated bright stars. The central dashed ellipse corresponds to the region within two half-light radii of the dwarf galaxy, assuming the structural parameters listed in Table~\ref{parameters}. \emph{Right panels:} Color-magnitude diagrams within two half-light radii of And~XXI (middle panel) and, for comparison, of a field region at a distance of $\sim20'$ covering the same area after correcting from gaps in the survey coverage (right-most panel). The galaxy's RGB is clearly visible as an overdensity of stars with $0.8\lta g-i\lta1.5$ and $i\gta21.2$ that does not appear in the reference CMD.}
\end{center}
\end{figure*}

\begin{figure*}
\begin{center}
\includegraphics[width=0.44\hsize,angle=270]{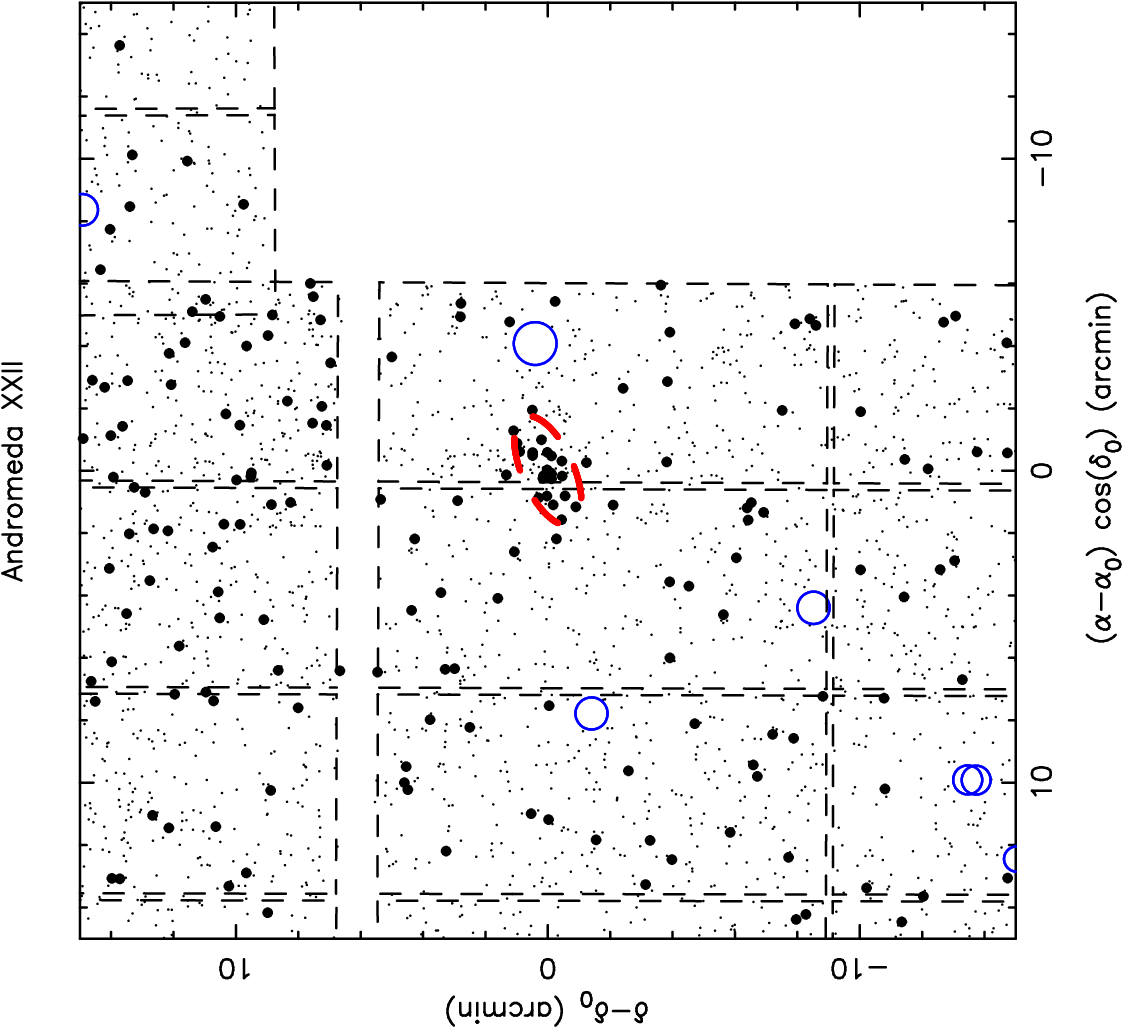}
\includegraphics[width=0.44\hsize,angle=270]{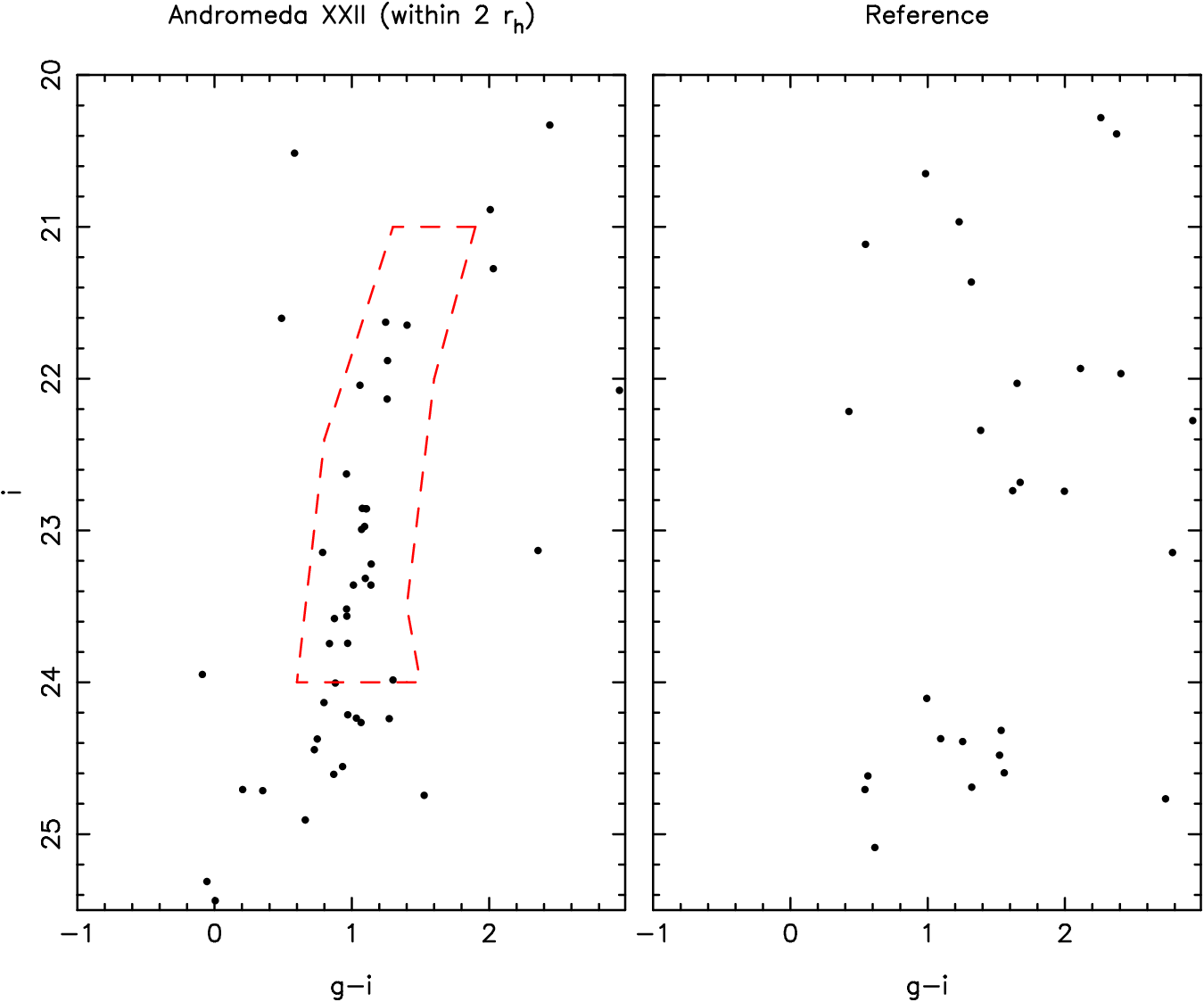}
\caption{\label{map_AndXXII} Same as Figure~\ref{map_AndXXI} but for And~XXII. Although this system is much fainter, it still appears as a spatial overdensity of stars (left panel) that are aligned along a RGB in the CMD (middle panel), a feature that does not appear in the reference CMD (right panel).}
\end{center}
\end{figure*}

Both systems appear as overdensities of stars on the sky, as is visible in the left panels of Figures~\ref{map_AndXXI} and~\ref{map_AndXXII}. These stars are also aligned along a RGB that would be at, or close to, the distance of M31 or M33. The CMDs within 2 half-light radii of the dwarfs (determined in \S~\ref{strParam} below) are shown in the middle panels of these figures and, when compared to the CMD of reference fields chosen in an annulus covering the same area at a distance of $\sim20'$ from the dwarfs' centers (right panels), indeed reveal an alignment of stars that follow the typical shape of a RGB. Isolating these stars enhances the contrast of the overdensity of stars on the sky (large symbols in the left panels).

And~XXI is typical of the relatively bright dwarf galaxies that we have found before (such as And~XV or And~XVI). The overdensity of stars in the CMD region $24.5<i<25.0$ and $g-i\sim1.0$ corresponds to its horizontal branch but the degrading star/galaxy separation at these magnitudes, as well as the rapidly dropping completeness, makes it difficult to extract any accurate information from this region of the CMD. With only $\sim20$\,stars visible along its RGB, And~XXII is one of the faintest M31 satellites found so far.

\begin{table}
\caption{Derived properties of the satellites}
\label{parameters}
\begin{tabular}{l|cc}
\hline\hline
 & And~XXI & And~XXII \\
$\alpha$ (J2000) & $23^{\rm h}54^{\rm m}47.7^{\rm s}\pm 1.0^{\rm s}$ & $01^{\rm h}27^{\rm m}40.0^{\rm s} \pm 0.6^{\rm s}$\\
$\delta$ (J2000) & $42\deg28'15''\pm15''$ & $28\deg05'25'' \pm6''$ \\
$(l,b)$ ($\deg$) & $(111.9,-19.2)$ & $(132.6,-34.1)$ \\
$E(B-V)^\mathrm{(a)}$ & 0.094 & 0.076 \\
$(m-M)_0$ & $24.67\pm0.13$ & [24.50] $\lta25.07^\mathrm{(b)}$ \\
$D$ (kpc) & $859\pm51$ & [794] $\lta1033^\mathrm{(b)}$ \\
$r_\mathrm{M31}$ (kpc) & $152\pm31$ & [224] $\lta 315$\\
$r_\mathrm{M33}$ (kpc) & --- & [42] $\lta 227$\\
$\langle\FeH\rangle$ & $-1.8$ & $[-1.8]$ $\gta-2.5$ \\
$r_h$ (arcmin) & $3.5\pm0.3$ & $0.94\pm0.10$ \\
$r_h$ (pc) & $875\pm127$ & [217] $\lta 282$ \\
PA (North to East) ($\deg$) & $110\pm15$ & $114\pm15$ \\
$\epsilon=1-b/a$ & $0.20\pm0.07$ & $0.56\pm0.11$ \\
$M_V$ & $-9.9\pm0.6$ & $[-6.5\pm0.8]$ $\gta -7.0\pm0.8$\\
$L_V$ ($\lsun$) & $7.8\pm1.9\times10^5$ & $[0.3\pm0.1\times10^5]$ $\lta 0.6\times10^5$\\
$\mu_{V,0}$ (mag/arsec$^2$) & $27.0\pm0.4$ & $26.7\pm0.6$\\
\end{tabular}\\
(a) \citet{schlegel98}\\
(b) Given the highly unreliable distance estimate of And~XXII from its TRGB, parameters between square brackets are determined assuming  the average distance of M31/M33. The limit on all parameters, determined from the distance limit measured from the brightest star along And~XXII's RGB, are also given in the same column.\\
\end{table}


All the parameters of the two dwarf galaxies determined in this section of the paper are summarized in Table~\ref{parameters}.

\subsection{Distances and metallicities}
\label{distances}
\begin{figure*}
\begin{center}
\includegraphics[width=0.30\hsize,angle=270]{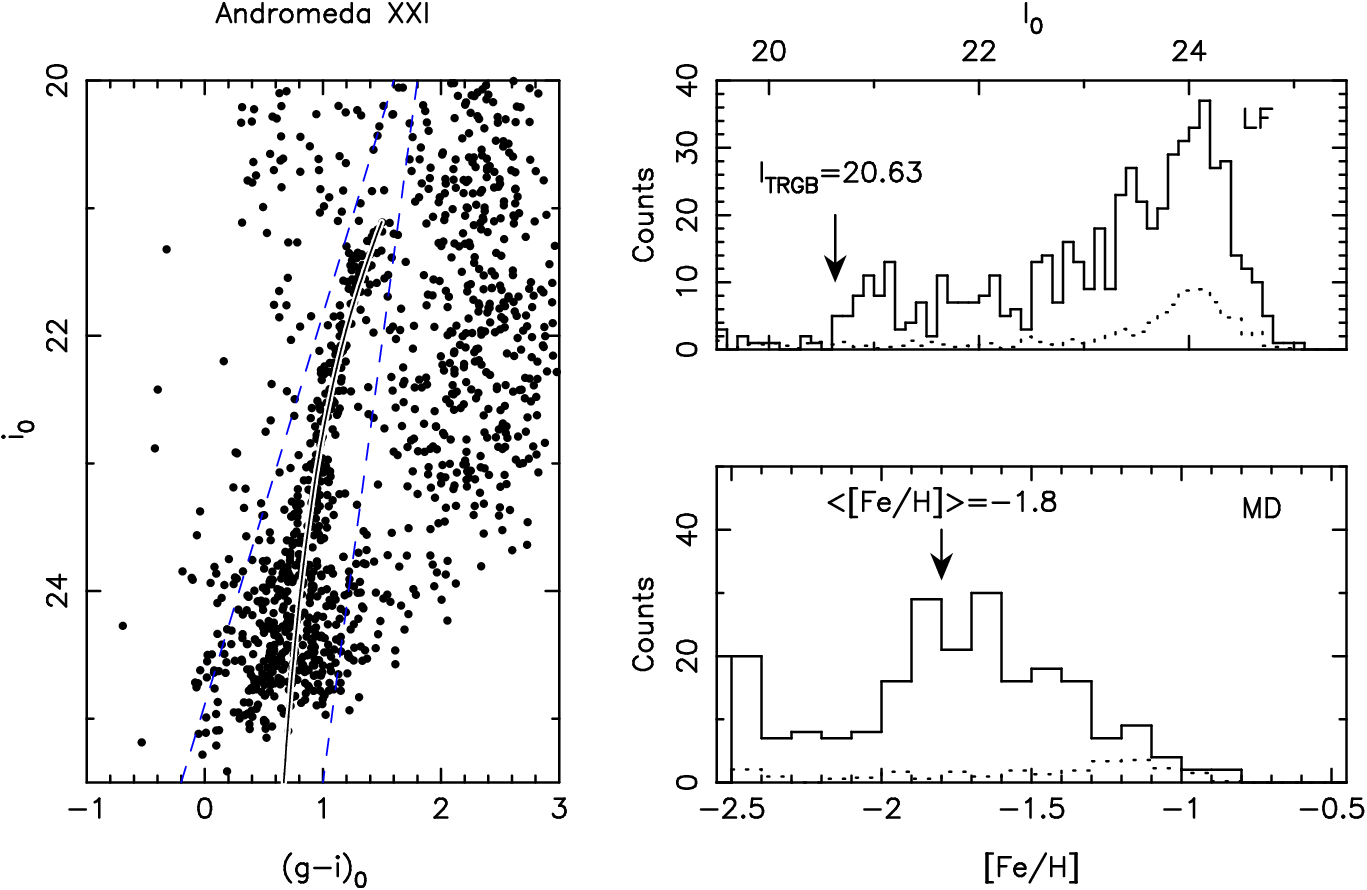}
\includegraphics[width=0.30\hsize,angle=270]{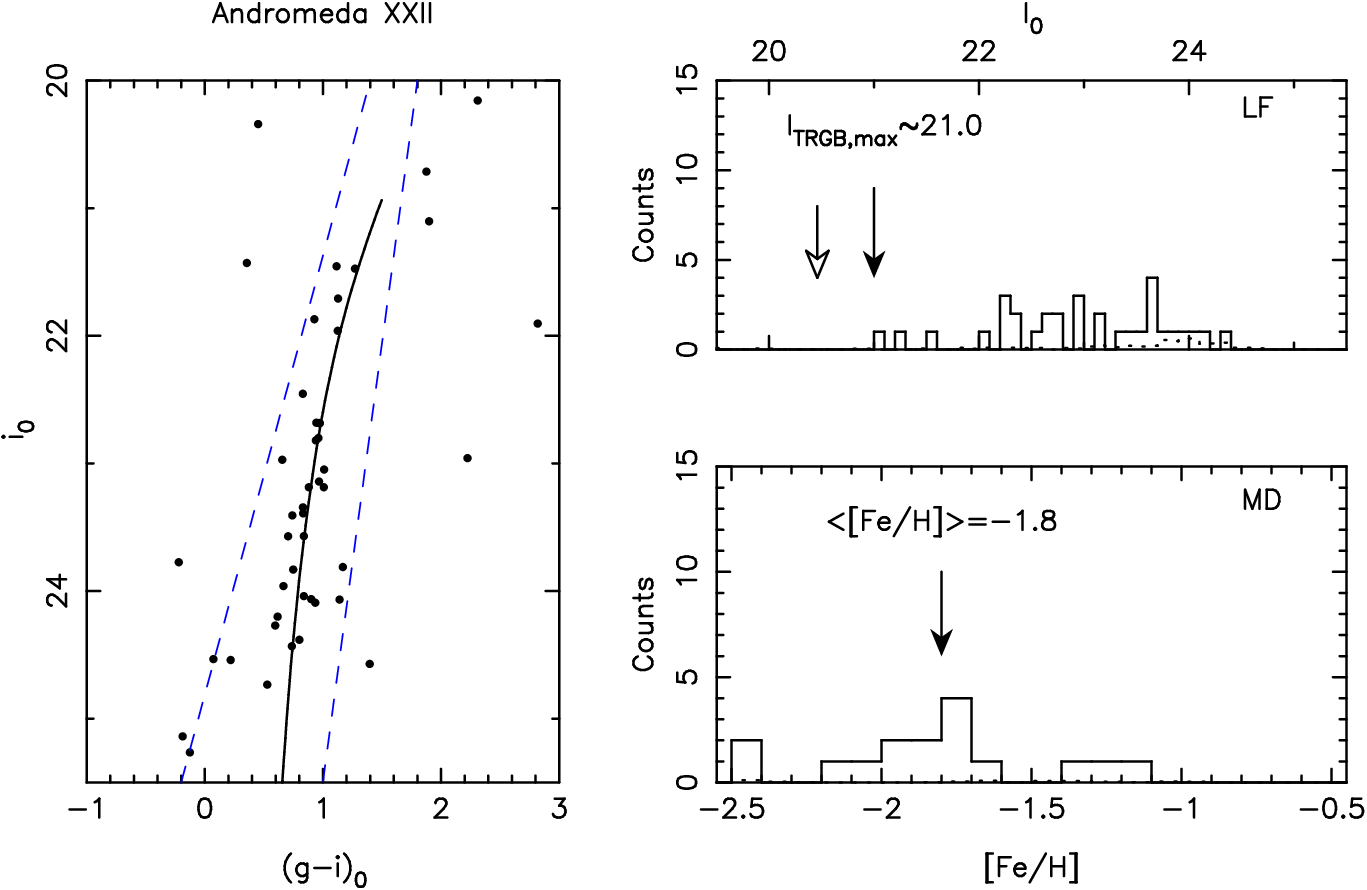}
\caption{\label{LF} \emph{Left panels:} The CMD (left), luminosity function (LF, top right) and metallicity distribution (MD, bottom right) of And~XXI's stars, located within two half-light radii and within the two dashed lines overlaid on the CMD. An additional cut to remove stars fainter than $i_0=23.5$ has also been applied to determine the MD. Both the LF and MD panels also show the scaled contribution of fore/background contamination as a dotted histogram, which is clearly negligible. The LF shows an increase of counts as one reaches the TRGB, which is estimated to be at $I_0=20.63$ and indicated by the arrow. This corresponds to a distance modulus of $(m-M)_0=24.67\pm0.13$. The MD is used to determine the average metallicity of the dwarf $\langle\FeH\rangle=-1.8$, also indicated by the arrow. The \citet{dotter08} 13\,Gyr isochrone with $[\alpha/\textrm{Fe}]=+0.0$ overlaid on the CMD corresponds to this metallicity and distance modulus and is in very good agreement with the satellite's RGB. \emph{Right panels:} Same as the left panels, but for And~XXII. In this case, the LF is so underpopulated that we indicate both the magnitude of the brightest star that is likely to be a RGB star ($I_0\simeq21.0$, filled arrow) and the average of the M31/M33 distance that we conservatively assume for And~XXII throughout the paper ($I_0=20.46$, hollow arrow).}
\end{center}
\end{figure*}

It is difficult to accurately estimate distances to dwarf galaxies without photometric observations that reach down to their horizontal branches. Indeed, the tip of the RGB (TRGB), which is often used as a distance indicator since its magnitude in the I-band only slightly varies with the properties of the dwarfs' stellar populations (e.g. \citealt{bellazzini01}), is not sampled well enough in faint systems that suffer from `CMD shot-noise': these systems do not contain enough stars for their CMD to be reliably sampled by their few stars, especially along their RGB \citep{martin08b}. In the case of And~XXI, this is not so much of an issue as it contains a reasonable number of RGB stars and the TRGB is therefore well defined. From the luminosity function (LF) of stars within two half-light radii of the dwarf's center (dashed ellipse on the map of Figure~\ref{map_AndXXI}) and within the two dashed lines of the dwarf's de-reddened CMD presented in Figure~\ref{LF}, we estimate $I_{0,\mathrm{TRGB}}=20.63\pm0.04$ by applying the algorithm of \citet{mcconnachie04}\footnote{The $gi$ MegaCam magnitudes were first converted to the INT $Vi'$ system following the color equations derived by \citet{ibata07} and then to the Landolt $VI$ system following the transformations of \citet{mcconnachie04}.}. The complete LF is shown on the top-right panels of the same figure. Along with the absolute TRGB magnitude of $M_I=-4.04\pm0.12$ \citep{bellazzini01}, we determine a distance modulus of $(m-M)_0=24.67\pm0.13$, or a distance of $D=859\pm51\kpc$.\footnote{The $g$-band LF of the dwarf, going deeper than the combined $g$- and $i$-band catalogue used to create the CMD of Figure~\ref{map_AndXXI}, also shows a peak consistent with that of And~XXI's horizontal branch at the measured distance modulus.}

In And~XXII, on the other hand, the brightest possible member RGB star that can be seen in the CMD of stars within two half-light radii is much less likely to be close to the TRGB and thus to yield a reliable distance. It can only provide an upper limit to the distance modulus of this dwarf. This star, with $I_0=21.03$ (see the top-right panel of Figure~\ref{LF}) converts to a maximum distance modulus of $(m-M)_{0,\mathrm{max}}=25.07$ or $D_\mathrm{max}=1033\kpc$. This limit to And~XXII's distance clearly makes this dwarf galaxy a likely member of the M31/M33 satellite system. The $g$-band LF of And~XXII also shows a tentative detection of its horizontal branch, at a distance of $\sim940\kpc$, very near the limit of the survey. Although this measurement is very uncertain, the absence of a clear horizontal branch at brighter magnitude would imply that And~XXII is at least as distant as And~XXI. For the moment, we will assume a distance to And~XXII that is the same as that to M31/M33 ($(m-M)_0\simeq24.50$ or $794\kpc$) until deeper photometric data provide a better constraint, keeping in mind that the dwarf is likely to be more distant than this.

Figure~\ref{LF} also presents the metallicity distribution (MD) of the stars that were used to construct the LF of the two satellites. The metallicity estimates of individual stars were obtained by comparison with the \citet{dotter08} set of $13\Gyr$ isochrones\footnote{These isochrones are available in the SDSS photometric system that is different, but very close, to the MegaCam one. The isochrones are therefore  recomputed for the MegaCam system from the color equations provided in the headers of the MegaCam fits images. See \url{http://www.cfht.hawaii.edu/Science/CFHTLS-DATA/megaprimecalibration.html} for more details.} with $[\alpha/\textrm{Fe}]=+0.0$\footnote{Using the larger alpha-enhancement of $[\alpha/\textrm{Fe}]=+0.4$ usually found in metal-poor dwarf galaxies (e.g. \citealt{tolstoy09}) yields \FeH\ values that are ~0.2\,dex more metal-poor but are a noticeably worse fits to the RGB of And~XXI.}, placed at the distance moduli determined above. Each star is given a metallicity determined by the ratio of the distances to its two closest isochrones. These have a metallicity ranging from $\FeH=-2.5$ to $\FeH=+0.5$, with steps of 0.1\,dex. There is no significant contamination of these MD from the field population (dotted histograms), except for the very few, most metal-rich stars to appear in And~XXI's MD. When this contamination is removed, both galaxies are determined to be quite metal-poor, with an average metallicity of $\langle\FeH\rangle=-1.8$. The corresponding isochrones, overlaid on the CMDs of the same figure, show good agreement with the location of the observed RGB stars that revealed the satellites. Of course, if the distance modulus of And~XXII is larger than the $(m-M)_0=24.50$ value adopted here, that would translate into a lower metallicity for this system, with a minimum $\langle\FeH\rangle=-2.5$ for the maximum distance obtained above. It should also be noted that, given the photometric uncertainties and the small color differences between metal-poor isochrones, the determination of photometric metallicities does not give a good handle on the metallicity spread in these systems and the width of the MDs should not be interpreted at face value.

\subsection{Structural parameters}
\label{strParam}
To determine the structural properties of the two dwarf galaxies, we apply the maximum likelihood technique developed by \citet{martin08b} that determines the best exponential profile (and background level) to fit the spatial distribution of a galaxy's stars. Compared to the original paper, the algorithm has been modified to account for incompleteness in the surroundings of the satellites, produced by gaps between the MegaCam CCDs, the survey edges, and the presence of saturated bright stars. The first effect is patently important in the case of And~XXI, which has a CCD gap running through its center. The best exponential profile is determined iteratively, after filling holes from the parameters of the best previous model. This step is performed by first binning the survey region around the dwarfs with small $0.1\times0.1\textrm{arcmin}^2$ pixels. Each pixel that falls over the edge of the survey, in a CCD gap, or behind a bright star is filled with a number of stars randomly generated from a Poisson distribution of mean the model density at this pixel's central position, computed from the best model from the previous iteration. These stars are then randomly distributed within the pixel. Convergence on the structural parameters model is usually achieved within five iterations but we proceed with eight iterations to secure convergence. Given the stochasticity introduced by filling the holes with artificial stars, we determine the contribution of the gap to the uncertainties on the model parameters via a Monte Carlo approach, performing 100 fits for each galaxy.

So as to limit the impact of background/foreground contamination, we apply the maximum likelihood fit to these stars that fall within the selection box overlaid on the CMDs of Figures~\ref{map_AndXXI} and~\ref{map_AndXXII}. These have a faint limit fixed at $i=24.0$ so that fainter regions where star/galaxy separation becomes an issue are avoided. The centroid ($\alpha_0$, $\delta_0$), half-light radius ($r_h$), ellipticity (defined as $\epsilon=1-b/a$ with $a$ and $b$ being, respectively, the major and minor axes scale lengths of the system) and position angle (PA) for which convergence was achieved are listed in Table~\ref{parameters}.

\begin{figure}
\begin{center}
\includegraphics[width=0.8\hsize,angle=270]{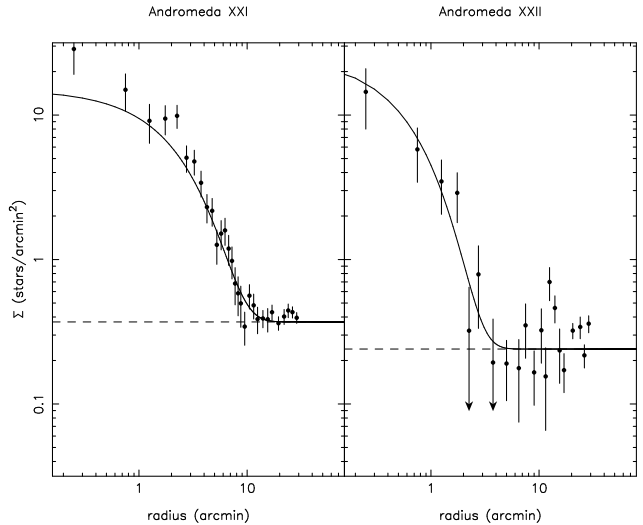}
\caption{\label{profiles} Stellar profiles of And~XXI and And~XXII. In both panels, the stellar density measured in fixed elliptical annuli from the PAndAS data is shown as dots, using the best ellipticity, position angle and centroid found by the fitting algorithm (the uncertainties are derived assuming Poisson statistics). These data points are corrected for incompleteness produced by CCD gaps, the edges of the survey and bright stars. The full lines represent the best models constructed from the best fit half-light radius and background level (the background level is represented by the thin dashed line). Although they are not fits to the data points, the profiles are in good agreement with them. Regions where some deviations are observed (at the center of And~XXI and in the outer part of And~XXII) are regions significantly affected by incompleteness.}
\end{center}
\end{figure}

In Figure~\ref{profiles}, the fitted exponential profiles are compared to the data, which are binned following the best centroid, ellipticity and position angle determined by the fit. The binned data points have been corrected for spatial incompleteness produced by CCD gaps, survey edges and bright stars. For both dwarfs, the agreement between profile and data is quite good. There are regions --- in particular in the inner parts of the And~XXI profile and the outer parts of the And~XXII one --- where deviations from the model are significant but these correspond to regions were a large completeness correction was applied.

\subsection{Magnitudes}
To measure the total magnitude of each dwarf galaxy, we first determine the flux contribution of all stars brighter than $i_0=23.5$ (where we know the survey to be complete\footnote{We have nevertheless checked the validity of our parameters with the fainter $i_0=24.0$ limit and find that, with this deeper limit, none of the derived parameters deviate by more than $\pm0.1$\,mag. from those listed in Table~1.}; \citealt{ibata07}) and within its half-light radius, following the best structural parameters determined above. We then correct this number for background/foreground contamination by counting the flux within the annulus between 16 and $20'$, scaled to the area covered within the half-light radius. Both these fluxes are also corrected for incompleteness coming from bright stars, CCD gaps and survey edges. For And~XXI and And~XXII, we measure partial magnitudes of $M_V=-7.7$ and $-4.3$ respectively. These numbers then need to be corrected to include the flux contribution of all stars fainter than $i_0=23.5$. In order to do this, we use the PandAS observations of another Andromeda satellite, And~III, that has also been observed in the survey under the same conditions as And~XXI and And~XXII. In the case of this satellite, we measure a difference of 2.2 magnitudes between the flux of stars brighter than the $i_0=23.5$ limit and the total magnitude of the dwarf $M_V=-10.2\pm0.3$ measured by \citet{mcconnachie06b}. This leads to total magnitudes of $M_V=-9.9$ and $-6.5$ for the two new dwarfs when we account for this difference. We estimate the uncertainties on these values to be $\pm0.6$ for And~XXI and $\pm0.8$ for And~XXII, stemming from the uncertainties on the structural parameters ($\pm0.2$), the distance estimates ($\pm0.1$ and $\pm0.3$), the CMD shot-noise of these sparsely populated dwarf galaxie ($\pm0.2$ and $\pm0.4$), and the total magnitude of And~III used for the calibration ($\pm0.3$) all added in quadrature, to which we add a systematic uncertainty of $\sim0.2$\,magnitudes to account for possible differences between the And~XXI and And~XXII luminosity functions and that of And~III.

From the total magnitude, we determine the surface brightness of the dwarfs, following equation (6) of \citet{martin08b} with the number of stars $N_*$ being replaced by the total luminosity of the dwarf, $L_V$, listed in Table~\ref{parameters}. This equation accounts for the structure of the dwarf and in particular its ellipticity, and yields central surface brightnesses of $\mu_{V,0}=27.0\pm0.4$ and $26.7\pm0.4\,\textrm{mag}/\textrm{arcsec}^2$ respectively for And~XXI and And~XXII. Their low surface brightnesses make them invisible on photographic plate and, contrary to the brighter And~XVIII \citep{mcconnachie08}, they do not appear on the Digitized Sky Survey.

\section{Discussion}

\subsection{Properties of the new dwarf galaxies}
The measured magnitude and surface brightness of both new dwarf galaxies are within the range of values measured for dwarf galaxies previously found and, as such, And~XXI and And~XXII are very typical Andromeda satellites. The structures of these two dwarf galaxies are also fairly typical of that of M31's satellites. Although the ellipticity of And~XXII is quite high, it is not the only elliptical dwarf spheroidal galaxy known to orbit Andromeda. And~III, for instance, has a similarly large ellipticity ($\epsilon=0.52\pm0.02$, \citealt{mcconnachie06b}).

With a half-light radius of almost $1\kpc$, And~XXI supports the size difference that has previously been noted to distinguish Andromeda from Milky Way satellites: an M31 dwarf galaxy is typically 2--3 times larger than its Galactic counterpart \citep{mcconnachie06b}. This distinction still holds for fainter dwarf galaxies like the ones recently discovered. And~XXI is the fourth largest dwarf spheroidal galaxy known in the Local Group after And~XIX ($r_h=1.7\pm0.1\kpc$; \citealt{mcconnachie08}), And~II ($r_h=1.12\pm0.02\kpc$; based on the exponential profile fit of \citet{mcconnachie06b} for consistency) and Sagittarius, which is clearly in the final throes of its tidal disruption by the Milky Way. It is also interesting to notice that two of these galaxies (And~XIX and And~XXI) are recent discoveries, hinting that some other large satellites could still elude us, especially given that we are biased against finding large faint galaxies, as they are more diffuse and therefore have a lower surface brightness. Galaxies as large and diffuse as And~XIX would also be very hard to find around the Milky Way but, on the contrary, it is unlikely that Milky Way satellites that share the properties of And~II and And~XXI ($r_h\sim1\kpc$, $M_V\lta-9.0$) could have been missed in the quarter of the Galactic sky covered by the SDSS \citep{koposov08,walsh09}. The largest Galactic satellite known to date was indeed discovered in the SDSS, but its half-light radius is only a mere $564\pm36\pc$. Such a distinction is puzzling and can be used, in conjunction with spectroscopic data, to provide strong constraints on the type of dark matter haloes that these satellites should inhabit \citep{penarrubia08a}.

\subsection{And~XXII: an M33 satellite?}
For And~XXII, the most intriguing question is whether it is a satellite of M33 or M31. M33 is a small spiral galaxy with a mass approximately one tenth the mass of M31. It resides at a distance of $204\kpc$ from the center of M31, presumably well-inside the M31 dark halo whereas measurements of its line-of-sight velocity and proper motion suggest that it is on a highly elongated orbit around M31.  The implication is that M33 is a satellite of M31 and likely passed much closer to M31 in the past than its present distance.  This association is supported by the existence of a strong warp in the M33 HI disk \citep{rogstad76} as well as the presence of a stellar stream extending from its stellar disk \citep{mcconnachie09}.

If And~XXII is shown to be a satellite of M33, then it would expand the known size of the M33 system by 50\%, from its recently discovered outermost globular cluster \citep{huxor09}. The existence of a satellite at this radius would also place a constraint on the strength of an M33--M31 interaction since such an interaction naturally strips material from the outer parts of M33. Finally, we point out that And XXII may be one of the few examples of a satellite's satellite, a phenomenon that naturally arises in the hierarchical clustering scenario of structure formation.

\begin{figure}
\begin{center}
\includegraphics[width=0.7\hsize]{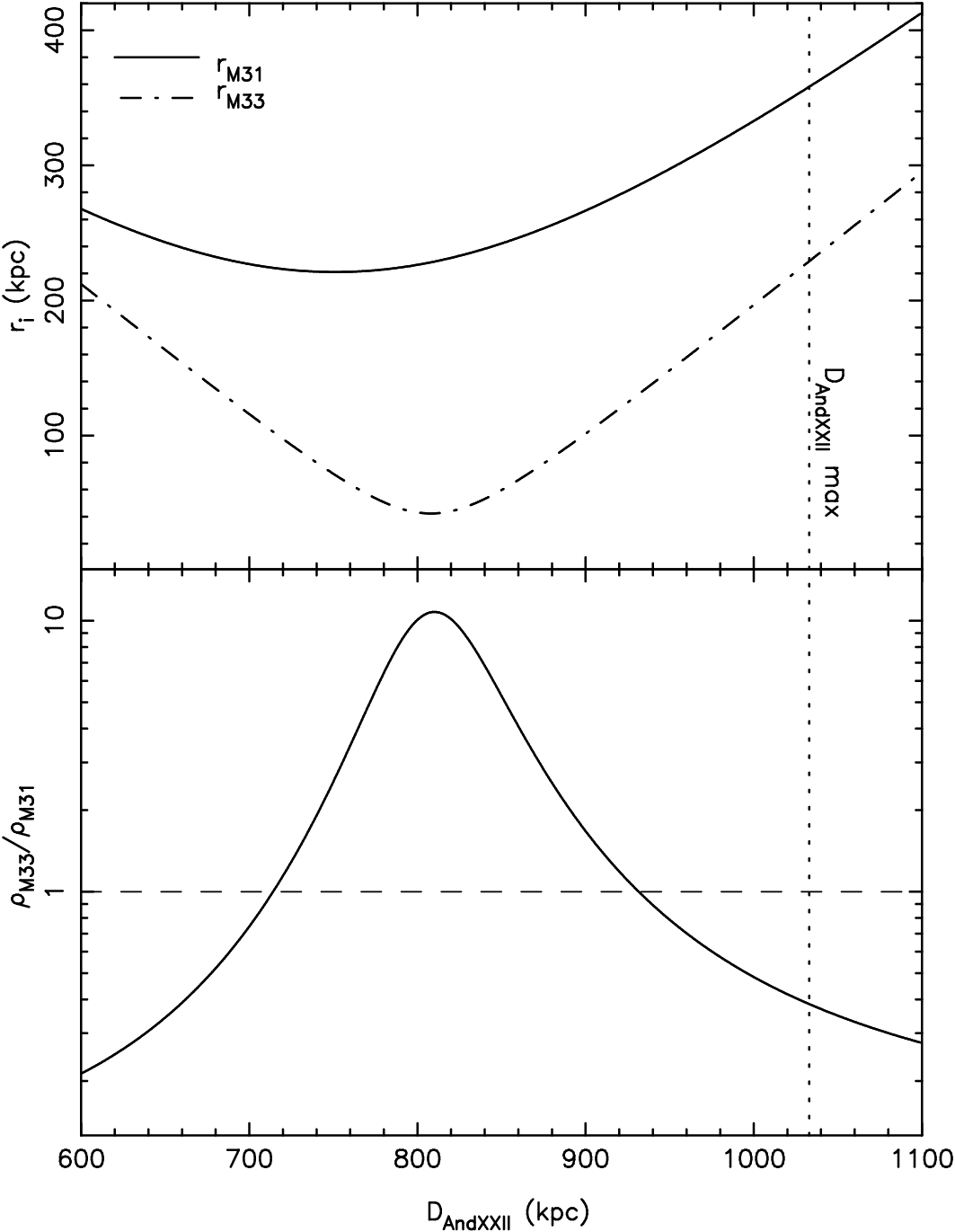}
\caption{\label{forces} \emph{Top panel:} The And~XXII/M31 (full line) and And~XXII/M33 distance (dash-dotted line) as a function of the And~XXII heliocentric distance. The maximum distance we estimate for And~XXII, $D_\mathrm{AndXXII}=1033\kpc$ is represented by the dotted line. \emph{Bottom panel:} The ratio of the M33's dark matter halo density to that of M31 as a function of the heliocentric distance to And~XXII. There is a reasonable distance range ($720\lta D_\mathrm{AndXXII}\lta930\kpc$) over which the density of the M33 dark matter halo dominates, hinting at a more likely association of And~XXII to Triangulum in this region.}
\end{center}
\end{figure}

And~XXII is, in projection, much closer to M33 than to M31. However, since the distance to And~XXII has not been determined, the true And~XXII/M31 and And~XXII/M33 distances are not known.  This is illustrated in the top panel of Figure~\ref{forces} that shows these distances as a function of the heliocentric distance to And~XXII, $D_{\rm AndXXII}$.

Cosmological simulations suggest that the density of satellites traces the dark matter density \citep{sales07a}. If true, the ratio of the M33 halo density to the M31 halo density at a given location provides a rough measure of the probability for a dwarf at this location to be a satellite of M33 relative to the probability that it is a satellite of M31. This ratio is shown in the bottom panel of Figure~\ref{forces}, as a function of the heliocentric distance to And~XXII. Here, we have assumed that the halos of M31 and M33 have NFW density profiles \citep{navarro97} defined by a virial mass, $M_{\rm vir}$, and a virial radius, $R_{\rm vir}$.  By definition, the mean density within the virial radius is a factor $\Delta$ times the background density where we take $\Delta\simeq 95$ \citep{maccio08}, as appropriate for WMAP5 values of the cosmological parameters \citep{komatsu09}. We assume $M_{\rm vir}=2\times10^{12}\msun$ for M31 and a mass that is ten times lower for M33, consistent with current observational constraints \citep{mcconnachie06c}. Concentration parameters are chosen as $c=10$ and $c=9$ respectively, following the relations of \citet{neto07} and \citet{maccio08}, which leads to virial radii of $R_{\rm vir}= 327\kpc$ for M31 and $R_{\rm vir} = 152\kpc$ for M33. Figure~\ref{forces} suggests that if the distance to And~XXII is found to be between $720\kpc$ and $930\kpc$, then it is more likely to be a satellite of M33 than of M31\footnote{It should be noted that recent modeling of the Andromeda gravitational potential suggests that it could be less massive and with a larger concentration than assumed here \citep{geehan06,seigar08}. This might expand, or otherwise narrow the range over which And~XXII is more likely to be an M33 satellite. However, given the lack of a similarly accurate model for M33, we currently favor cosmologically-motivated parameters for both galaxies, while we note that better models are in the process of being developed, partially constrained by PAndAS data.}.

An alternative argument can be made based on the notion of a Hill sphere which roughly defines the region where M33 rather than M31 dominates the dynamics of small systems.  In the limit of circular orbits (admittedly, a bad assumption for this problem) the radius of the Hill sphere for M33 is given by $r_{\rm Hill} = D_{\rm M33-M31} (M_{\rm M33}/3M_{\rm M31})^{1/3}$. The range of distances where And~XXII is likely to be an M33 satellite is then narrower ($757<D_{\rm AndXXII}<856\kpc$).

While our arguments are admittedly simplistic, they illustrate the importance an accurate distance measurement for And XXII will have in settling this issue.  A measurement of the line-of-sight velocity for And~XXII will also provide valuable information as will detailed modelling of the M31--M33 interaction.

\section{Conclusion}
In this paper, we present the discovery of two Local Group dwarf galaxies from the first year data of the Pan-Andromeda Archaeological Survey (PAndAS). Combined with previous installments of the MegaCam survey \citep{ibata07}, the M31 and M33 surroundings have been mapped over $\sim220$ deg$^2$ and unveiled 10 new dwarf galaxies. The two new satellites, Andromeda~XXI and Andromeda~XXII, have properties that are typical of other Andromeda satellites:

\begin{itemize}
\item Andromeda~XXI is a reasonably bright galaxy, with $M_V=-9.9\pm0.6$, located slightly farther than M31 at $859\pm51\kpc$, or at $152\pm31\kpc$ from Andromeda itself and is metal-poor ($\langle\FeH\rangle=-1.8$, assuming $[\alpha/\textrm{Fe}]=+0.0$). Its size is its most extraordinary property and, with a half-light radius of $875\pm127\pc$, it is the third largest dwarf spheroidal galaxy of the Local Group, emphasizing the greater size of M31 satellites with respect to Milky Way satellites.
\item Andromeda~XXII is a much fainter system, with a total magnitude of only $M_V=-6.5\pm0.8$; it is also metal-poor with $\langle\FeH\rangle=-1.8$ (also under the assumption of $[\alpha/\textrm{Fe}]=+0.0$. Its physical properties are made difficult to constrain by the absence of a good distance estimate originating from the low number of stars that populate its red giant branch. An accurate distance estimate will have to rely on the magnitude of And~XXII's horizontal branch from deeper photometry. We nevertheless raise the interesting possibility that, given its proximity to M33, at least in projection, And~XXII could be the first ever discovered Triangulum satellite. We show that the dwarf galaxy has a reasonably large heliocentric distance range ($720\lta D_\mathrm{AndXXII}\lta930\kpc$) over which it would more probably be a satellite of M33 than of M31.
\end{itemize}

\appendix
\section{Reflections on dwarf galaxy naming conventions}
\label{naming_conventions}
Naming conventions of Local Group dwarf galaxies were developed somewhat empirically with the increasing number of dwarf galaxies discovered over the last century. It is commonly agreed that Milky Way satellite dwarf galaxies should be named by the name of the constellation they are located in. For practical reasons, this rule has sometimes been extended to Local Group satellites that are not orbiting the Milky Way, such as, for instance, Sextans~A or Sextans~B, but breaks down in the vicinity of Andromeda. There, it seems more practical to name Andromeda satellites with the Andromeda prefix to emphasize that they could belong to the M31 satellite system. This is the reason why the Pegasus dwarf galaxy \citep{karachentsev99} is also usually called Andromeda~VI \citep{armandroff99} or why the Cassiopeia dwarf galaxy \citep{tikhonov99} is also called Andromeda~VII. But this convention is not without difficulties. Indeed, there are Local Group satellites that happen to be in the Andromeda constellation but are probably not satellites of Andromeda. Andromeda~XII, with its large radial velocity, might not be orbiting M31 \citep{chapman07} and Andromeda~XVIII is so far away behind M31 that it is also very unlikely to be orbiting M31 itself. Finally and contrary to what was initially proposed, And~IV is a background galaxy that is completely unrelated to M31 \citep{vandenbergh72,ferguson00}, whereas the overdensity of planetary nebulae dubbed And~VIII was later found to be made of normal disk planetary nebulae \citep{morrison03,merrett06}.

The situation becomes even more confusing with the satellite that we have dubbed Andromeda~XXII. Although its heliocentric distance makes it a likely satellite of M31, it resides in the Pisces constellation, and its proximity to M33 makes it a possible satellite candidate of the Triangulum galaxy. Without any knowledge of its kinematics (and obtaining radial velocity measurements for its brightest RGB stars will be trying at best, given their faintness; see Figure~\ref{map_AndXXII}), it is a priori impossible to decide on whether to call this satellite Andromeda~XXII, Triangulum~I, or whether it should take its name from the Pisces constellation. To make matters even worse, there is already an M31 dwarf galaxy, LGS~3, that is also sometimes called the Pisces dwarf and it happens that four `Andromeda satellites' (And~II, And~XIII, And~XIV and And~XVI) already lie in the Pisces constellation.

\begin{table}
\caption{Alternative names of `Andromeda satellites'}
\label{alternative_names}
\begin{tabular}{l|cc}
\hline\hline
Usual name & Alternative name\\
LGS~3 & Pisces~(I)\\
Andromeda~II & Pisces~II\\
Andromeda~VI & Pegasus\\
Andromeda~VII & Cassiopeia\\
Andromeda~XIII & Pisces~III\\
Andromeda~XIV & Pisces~IV\\
Andromeda~XVI & Pisces~V\\
Andromeda~XXII & Pisces~VI, (Triangulum~I)$^\mathrm{(a)}$\\
\end{tabular}\\
(a) This alternative naming shall supersede the `Andromeda~XXII' name if this satellite is confirmed to be a Triangulum satellite instead of an Andromeda satellite.\\
\end{table}

To avoid further confusion, we suggest to follow the dual naming introduced with the discovery of the Pegasus/And~VI and Cassiopeia/And~VII dwarf galaxies as listed in Table~\ref{alternative_names}.

\bibliographystyle{apj}

\clearpage

\clearpage

\end{document}